# Negative Impact and Probable Management Policy of E-Waste in Bangladesh

S. M. Rezaul Karim, Shariful Islam Sharif, Md. Anisur Rahman Anik

Department of Electrical and Electronics Engineering

IUBAT-International University of Business Agriculture and Technology

Dhaka-1230, Bangladesh

**rkarim@iubat.edu, sisharif41@gmail.com, anik4161@gmail.com**

**ABSTRACT–** Due to the wider usage of electrical and electronic products throughout the country, Electronic waste (E-waste) is growing and creating problems. E-waste contains toxic materials which may cause severe health risk and environmental damage. Bangladesh produces almost 2.7 million metric tons of E-waste per year and stays on high risk. This paper indicates the problem and suggests proper management of E-waste.

**Keywords—** E-waste, toxic materials, impact, e-waste management, recycle.

## I. INTRODUCTION

"Digital Bangladesh" vision 2021 aims to make our nation stronger through effective use of modern technology on important sectors and areas like education, health, communication and alleviation of poverty levels. The more digital devices are used the more E-waste will be produced. Electronic waste (E-waste) is a term for electronic products that have become unwanted, non- working or obsolete and have essentially reached the end of their useful life. E-waste is created from anything electronics: computers, TVs, monitors, cell-phones, CD, DVD, PDAs, fax machines, printers etc (E-Waste Guide Info). In the current era of digital revolution, we can see and feel the presence of electronics goods everywhere around us. E-waste contains more than 1000 different toxic materials such as lead, mercury, copper, cadmium, beryllium, barium etc (Ahmed, 2011). This may cause severe health risk and environmental damage. So, we need to recycle all effective and worthy materials from e-waste. Bangladesh is a developing country which is suffering from the rapid growing problems of e-waste. There are lack of awareness and information gap among people about e-waste problem in Bangladesh.

## II. CURRENT SCENARIO OF E-WASTE IN BANGLADESH

From BTRC (Bangladesh Telecommunication Regulatory Commission) data, it is evident that there are 129.584 million subscribers of mobile phones are available now (BTRC, 2017). The lifespan of a mobile phone currently stands at a maximum of 2.5 years or less in some cases. It indicates that these used mobile phones will be turned into waste within the next two or three years. As per G & R's data, there are 19 million users who are using desktop or laptop now and its lifespan 3-5 years (G & R).

**Table-01: E-Producer Lifespan (HG)**

| E-Product | Average Lifespan (Years) |
|---|---|
| Mobile | 1.5-2.5 |
| Computer | 3-5 |
| Television | 9-14 |
| Refrigerator | 15-17 |

Since our country is becoming digitalized in the near future, the usage of electronic equipment will increase at a faster pace. Some E-waste is being collected, separated, demolished and recycled in the informal sector based in Dhaka's urban slums but most are released in the canal, rivers, drains, lake, open space and landfills. The total amount of produced e-waste in Bangladesh is 2.7 million metric tons per year (Alam and Bahauddin, 2015). Reasons behind this increasing are unconsciousness and absence of law regarding e-waste. If e-waste is not recycled and disposed of in environmentally prudent manners, it can pose serious threats to both human health and the environment.

**Figure 1: Awareness about E-Waste (RTV-News)**

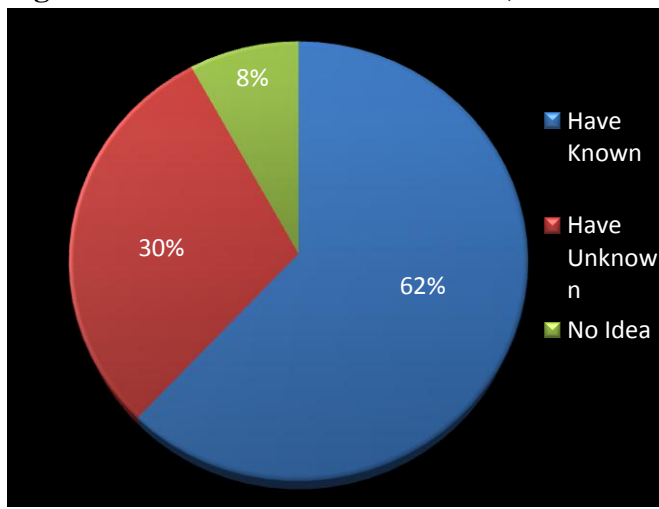

In Bangladesh each year more than 15% child worker died during and after effect of e-waste recycling and more than 83% are blooming by toxics substances and become unwell and live with long term illness. According to ESDO approximately fifty thousand children's are involved in the non-formal e-waste collection and recycling process and 40% are involved in ship breaking carapace (Islam, 2016).

### III. WORLDWIDE CURRENT SCENARIO OF E-WASTE

In 2017, the amount of e-waste is approximately 72 billion tons worldwide. USA is the largest producer of e-waste in the world. Indeed, electronic waste is a global problem requiring a global solution. The report by the United Nations University shows, per year 3.7 and 15.6 kg e-waste (per person) are being produced in Asia and Europe respectively (The Guardian, 2017).

**Figure 2: Top 10 E-waste Producer Countries (China Daily)**

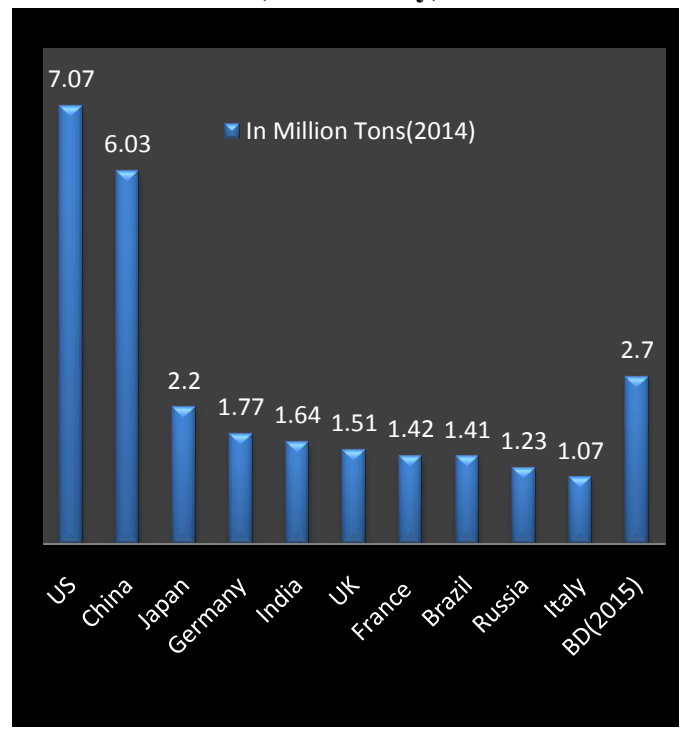

In 1990s, government in the EU, Japan, USA and some others industrialized countries began to tighten the regulatory framework against e-waste. They also set up e-waste retrieval, recycling systems (Alam and Khalid, 2015).

### IV. NEGATIVE IMPACT ON HUMAN HEALTH AND ENVIRONMENT

When we look at a computer or cell phone, it does not seem dangerous. Typically only the outer casing is visible but it's what's inside that poses a threat to the environment, people and animals. Electronics products contain lead, mercury, copper etc which creates a variety of problem shown below:

### Table-02: Effect of E-waste on Human Health (Krishna and Saha)

| Hazardous Material | Effect of E-Waste |
|---|---|
| Arsenic | Can affect skin and can decrease nerve conduction velocity. Chronic exposure to arsenic may cause lung cancer and sometimes be fatal. |
| Barium | Can affect heart muscle. |
| Beryllium | May cause lung diseases. |
| Chromium | Can damage liver, kidneys and may cause asthmatic bronchitis and lung cancer. |
| Cadmium | May cause severe pain in the joints and spine. It affects the kidneys and softens bones. |
| CFC | May affect the ozone layer. It may cause skin cancer in human and genetic damage in organisms. |
| Dioxin | These are highly toxic to animals and can lead to malfunction of fetus, decreased reproduction and growth rates, affect immune system. |
| Mercury | Affects the central nervous system, kidneys and immune system it impairs fetus growth. May cause brain or liver damage. |
| PCB | May cause cancer in animals, can affect the immune system, reproductive system, nervous system, endocrine system. PCBs persistently contaminate in the environment and cause severe damage. |
| Lead | May affect kidneys, reproductive systems, nervous, connections. May cause blood and brain disorders, sometimes may be fatal. |
| PVC | PVC contains up to 56% chlorine and when burnt, produces hydrogen chloride gas which intern produces hydrochloric acid that is dangerous to respiratory system. |

### Table-03: Effect on Environment (Wikipedia, E-Waste)

| E-Waste Component | Process Used | Potential Environmental Hazard |
|---|---|---|
| Cathode ray tubes (used in TVs, computer monitors, ATM, video cameras and more) | Breaking and removal of yoke, then dumping | Lead, barium and other heavy metals leaching into the ground water and release of toxic phosphor |
| Printed circuit board (image behind table-a thin plate on which chips and other electronic components are placed) | De-soldering and removal of computer chips; open burning and acid baths to remove metals after chips are removed. | Air emissions and discharge into rivers of glass dust, tin, lead, brominates dioxin,, beryllium cadmium and mercury |
| Chips and other gold plated components | Chemical stripping using nitric and hydrochloric acid and burning of chips | PAHs, heavy metals, brominates flame retardants discharged directly into rivers acidifying fish and flora. Tin and lead contamination of surface and ground water. Air emission of brominates dioxins, heavy metals and PAHs |
| Plastics from printers, keyboard, monitors, etc | Shredding and low temp melting to be reused | Emission of brominates dioxins, heavy metals and hydrocarbons. |
| Computer wires | Open burning and striping to remove copper | PAHs released into air, water and soil. |

### V. DISADVANTAGES OF E-WASTE

- Arsenic, barium, beryllium, chromium and cadmium affect on skin, heart, lung and kidney. So, damage of these important organs maybe the reason to reduce average longevity of our people.
- CFC affects on genetic system. If CFC is increased at a high rate then our genetic system will be reached in a bad condition which is a threat for us.
- Dioxin damages the production, reproduction, and growth rate of fetus for animal. If this occurs then the birth rate of animal will be decreased and it will damage the whole eco-system which is very unexpected to us.
- Throwing of e-waste in the land is one of the reasons for land pollution and this pollution is responsible for sterile land. Since our land is cultivable and we are cultivating it for producing our food. So, become sterilized of our land is menacing.
- Different toxic e-waste materials are directly thrown into water and that creates water pollution. Due to water pollution, fish and other insects which live in water are also facing trouble. If this pollution continues then the extinction of water insects and fish is certain.

### VI. PROPOSED IDEA FOR E-WASTE MANAGEMENT

- Companies deal in electrics devices can offer their customers to submit expired / damaged products to the respective E-waste point/customer care to avail discount in buying next product from them. Then the companies will recycle those devices. Government donation and support should be provided to the companies.

- Government can impose law to the concerned companies for setting e-waste points and following a common guideline of e-waste management. There can be a centralized coordination system and recycle center in the country.
- Disobeying companies can be punished by fining and canceling the license.
- Segregation of E-waste from regular waste must be done at time of collecting waste in metropolitan area. Setting e-waste bins beside dustbins will help the segregation process.

## VII. ADVANTAGES OF E-WASTE MANAGEMENT

- Reduces the amount of waste sent to landfills and combustion facilities.
- Prevents pollution by reducing the need to collect new raw materials.
- Saves energy.
- Reduces greenhouse gas emissions that contribute to global climate change.
- Helps sustain the environment for future generations.

**Figure-3: Proposed Life Cycle of E-waste**

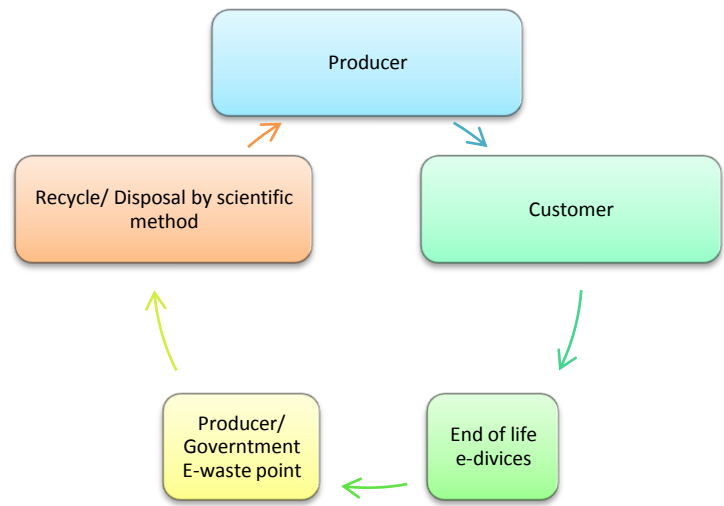

## VIII. PROPOSED METHOD FOR E-WASTE RECYCLING

At present the most efficient and scientific method of e-waste management is EXIGO recycling process. EXIGO is an Indian e-waste management company. We may follow EXIGO recycling method.

**Collection:** E-waste will be collected from various e-waste points every week.

**Transportation:** The collected e-waste is transported to the centralized recycle center in safe and secure manner using a closed container vehicle as per as government norms.

**Segregation**: Upon unloading, electronics waste segregation is done based on the size and available of components at the factory premises.

**Dismantling**: After segregation of e-waste components dismantle separately.

**Recycling**: After collecting all important materials from e-waste then do recycling and dispose the remaining hazardous waste through TSDF (EXIGO Recycling).

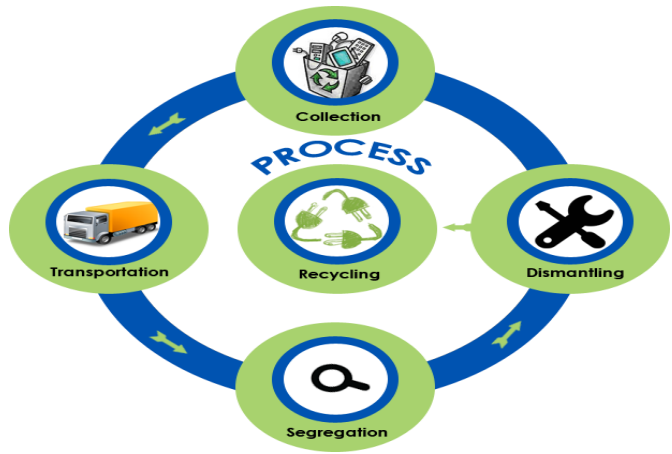

**Figure-4: E-waste recycling process by EXIGO**

## IX. CONCLUSION

It is time to address E-waste problem globally. A global solution is very much required here. As Bangladesh is a densely populated country the degree of impact on environment and population will be high. Awareness creation can be the first point to reduce its bad impact.